\documentclass[10pt,citeautoscript,floatfix,aps,prb,twocolumn,
      superscriptaddress]{revtex4-2}

\usepackage{amsmath}
\usepackage{physics}
\usepackage{appendix}
\usepackage{graphicx}
\usepackage{epstopdf}
\usepackage{natbib}
\usepackage{multirow}
\usepackage{array}
\usepackage[usenames,dvipsnames]{color}
\usepackage{dcolumn}
\usepackage{bm}
\usepackage{amssymb} 
\usepackage{booktabs}  
\usepackage{siunitx}

\usepackage{soul}  
\usepackage{changes}

\usepackage[colorlinks=true,linkcolor=blue,citecolor=blue,urlcolor=magenta]{hyperref}

\makeatletter
\let\Hy@footnote@currentHref\relax
\makeatother

\begin{document}
\title{Spin-orbit coupling renormalization of the natural optical activity of 
	Pb$_5$Ge$_3$O$_{11}$ from first-principles}
\author{Asier Zabalo}
\affiliation{Physique Théorique des Matériaux, Q-MAT, CESAM, Université de Liège, B-4000 Sart-Tilman, Belgium}
\author{Massimiliano Stengel}
\affiliation{Institut de Ciència de Materials de Barcelona (ICMAB-CSIC), Campus UAB, 08193 Bellaterra, Spain}
\affiliation{ICREA-Institució Catalana de Recerca i Estudis Avançats, 08010 Barcelona, Spain}
\author{Eric Bousquet}
\affiliation{Physique Théorique des Matériaux, Q-MAT, CESAM, Université de Liège, B-4000 Sart-Tilman, Belgium}
\date{\today}
\begin{abstract}
We present a first-principles study of the natural optical activity of the gyroelectric Pb$_5$Ge$_3$O$_{11}$ crystal, explicitly accounting for spin-orbit coupling (SOC) effects.
We derive a new analytical expression for the gyration coefficients within the recent framework of 
long-wavelength density-functional perturbation theory [Phys. Rev. Lett. \textbf{131}, 086902 (2023)], which significantly improves computational efficiency by reducing the number of required response functions and includes spin-orbit coupling effects.
We use this implementation to investigate 
the evolution of Pb$_5$Ge$_3$O$_{11}$'s optical rotation
across the ferroelectric double-well, from the paraelectric $P\bar{6}$ phase to the ferroelectric $P3$ phase. 
Our results demonstrate that, in addition to the substantial renormalization 
of the double-well energy, spin-orbit coupling contributions play an equally 
crucial role in the natural optical activity, largely through purely electronic 
contributions, while SOC-induced structural relaxation effects are minor.
\end{abstract}

\maketitle

\section{Introduction}
Lead germanate (Pb$_5$Ge$_3$O$_{11}$), referred to as PGO throughout this text, was first grown back
in 1971 \cite{10.1063/1.1653487,doi:10.1143/JPSJ.31.616}.
It is known to undergo a ferroelectric transition at $450$ K, 
bringing the system from a paraelectric (PE)
phase with the $P\bar{6}$ space group to a ferroelectric (FE) phase with the $P3$ space group.
The latter belongs to a non-enantiomorphic Sohncke space group, 
which is compatible with chirality,
meaning that left- and right-handed structures can crystallize in the same space group~\cite{bousquet2025structural}.
This $P\bar{6}$ to $P3$ structural phase transition has little impact on the lattice parameters of the crystal and, 
in both phases, 
the system can be described by a
primitive unit cell of $57$ atoms.
Since its synthesis
and the experimental confirmation of a stable ferroelectric phase, it
has been the subject of numerous experiments concerning 
its geometrical structure \cite{ivanov2022new,Kay01011975},
ferreolectricity \cite{fava2024} and ferroelectric domain walls~\cite{conroy2024, bak2020-PGOdomains, tikhonov2022-PGO_topology}, 
piezoelectricity \cite{10.1063/1.1661278}, 
dielectric response \cite{10.1063/1.1401092}, optical and electro-optical
\cite{VlokhLazkoShopa+1981+371+378,Adamenko_2008} properties.
In addition to extensive experimental studies, 
first-principles calculations 
based on density functional theory (DFT) have recently
been employed \cite{fava2024,PhysRevB.108.L201112} to shed light on
the microscopic origins of its structural phase transition, 
ferroelectricity and chirality, 
revealing that spin-orbit coupling (SOC)
effects play a critical role.
One of PGO's most remarkable properties is the possibility to switch its natural optical activity under an applied electric field, a property also called gyroelectricity~\cite{aizu1964, kizel1975, Konak1978} or ferrogyrotropy~\cite{wadhawan1979, wadhawan1982}. It is also worth noting that 
PGO exhibits an electrogyration coefficient
\cite{VlokhLazkoShopa+1981+371+378,https://doi.org/10.1002/adom.202500364}
among the largest ever recorded 
(reaching a value of 
$\gamma_{33}=(3.1\pm 0.3)\times 10^{-11}$ m/V in Cr-doped conditions).
Even in the absence of external electric fields, it displays 
spontaneous
natural optical activity (NOA), which
describes one of
the fundamental interactions between light and matter. 
Mathematically, natural optical
activity
arises in the constitutive relations as
\cite{landau1984electrodynamics} the first-order spatial dispersion
of the dielectric permittivity.
It primarily manifests in 
experiments via
optical rotation (OR), which quantifies the rotation of the plane of polarization of linearly polarized light as it passes through a crystal.

Even though nowadays experimental studies  \cite{kovnak1978induced,Iwasaki01021972}
about the 
natural optical activity in 
PGO are abundant 
(including investigations about 
the sign reversal 
of its optical rotation \cite{Iwasaki01021972} under an applied electric field, 
which stems from a change of structural handedness caused by the reversal 
of its polar order parameter~\cite{fava2024})
there has been little 
progress
from the theoretical 
forefront. 
This might be attributed to the absence, at least until quite recently,
of an accurate and accessible first-principles methodology for
the computation of NOA in crystals.
First-principles calculations are crucial for
understanding the microscopic origins of NOA and its interaction
with other material characteristics, such as electronic properties 
or structural chirality~\cite{bousquet2025structural}.
In recent years, several complementary approaches have been developed, 
including numerical methods based on the long-wavelength expansion of the electromagnetic vector potential response \cite{PhysRevLett.69.379,PhysRevB.48.1384}, 
analytical expressions for the NOA derived within the context of multipole theory of optical 
spatial dispersion in crystals
\cite{PhysRevB.82.245118,10.21468/SciPostPhys.14.5.118,PhysRevB.107.045201},
and Wannier interpolation techniques \cite{Urru_Wannier}. 
Notably, fully 
\textit{ab-initio} implementations have only recently been developed. 
One such approach, formulated within the GW-Bethe-Salpeter-Equation (GW-BSE) framework, 
was recently applied to the NOA of $\alpha$-quartz \cite{cvjj-kbck}, 
capturing excitonic effects at significant computational expense. 
A more efficient alternative had previously been developed
within the context of long-wave density-functional
perturbation theory (DFPT), as described in Ref. \cite{PhysRevLett.131.086902}, which is implemented 
in the open source \textsc{abinit} \cite{10.1063/5.0288278,GONZE2020107042}
package. 
Owing to its accuracy and computational efficiency, 
the latter implementation ---which completely avoids summations over empty states, 
does not require any numerical derivatives, and correctly treats 
self-consistent field contributions to the NOA--- emerges as the most compelling
choice in many cases. 
This methodology has been successfully applied to different
chiral crystals (including trigonal Se, $\alpha$-SiO$_2$ and $\alpha$-HgS)
and molecules.
This implementation, however, systematically neglects spin-orbit coupling
effects, which might be unjustified in presence of heavy elements like Pb or Bi,
thereby compromising the method's quantitative and qualitative predictive power in these specific cases.

In this work, we present a detailed first-principles study of the natural optical activity of the Pb$_5$Ge$_3$O$_{11}$ crystal.
We have extended the capabilities of the existing long-wave DFPT implementation
for the calculation of the natural optical activity
available in the \textsc{abinit} code in order to allow the treatment 
of spin-orbit coupling.
In addition, by exploiting the inherent electromagnetic gauge freedom of the theory, 
as discussed in Ref. \cite{PhysRevLett.131.086902},
we have derived a new analytical expression for the gyration coefficients. 
Our new implementation reduces the number of required response functions compared to the existing approach, thereby speeding up the calculation.
As a relevant case, we investigate the evolution of the natural optical 
activity in PGO as a function of the ferroelectric distortion from the $P\bar{6}$ to the $P3$ phase,
and evaluate the influence of spin-orbit coupling effects on the NOA.

The paper is organized as follows. 
In Sec. \ref{Sec_formalism}, after introducing some basic definitions, we 
summarize the implementation of the NOA as it
is currently available in \textsc{abinit}. 
We then extend the formalism to incorporate spin-orbit coupling
and present a new analytical expression for the NOA. 
The application of our implementation to the PGO crystal
is presented in Sec. \ref{Sec_results}. 
We finish the paper with the conclusions in Sec. \ref{conclusions}.

\section{Formalism}\label{Sec_formalism}
\subsection{Preliminary considerations}
We restrict our analysis to time-reversal symmetric insulators with broken inversion symmetry. Under such conditions, the natural optical
activity
\cite{agranovich1984crystal,10.1063/1.433207,10.21468/SciPostPhys.14.5.118} emerges as the 
only contribution to the first-order spatial dispersion of the 
dielectric permittivity tensor, 
$\epsilon_{\alpha\beta}(\omega,\mathbf{q})$, where $\omega$ is the frequency and $\mathbf{q}$ the wave vector. 
Expanding $\epsilon_{\alpha\beta}(\omega,\mathbf{q})$ in powers of 
$\mathbf{q}$, around $\mathbf{q=0}$, leads to 
\begin{equation}
	\epsilon_{\alpha\beta}(\omega,\mathbf{q})\simeq 
	\epsilon_{\alpha\beta}(\omega,\mathbf{q=0})+
	i q_\gamma \eta_{\alpha\beta\gamma}(\omega),
\end{equation} 
where $\eta_{\alpha\beta\gamma}(\omega)$ is the natural optical activity 
\cite{landau1984electrodynamics}, satisfying 
$\eta_{\alpha\beta\gamma}(\omega)=-\eta_{\beta\alpha\gamma}(\omega)$.
(Summation over repeated Cartesian indices is implicitly assumed throughout the whole text.)
To eliminate the redundancy in 
$\eta_{\alpha\beta\gamma}(\omega)$, which contains only 9 independent components out of 27, we contract it with the Levi-Civita symbol, $\varepsilon_{\gamma\delta\alpha}$, to obtain a more compact rank-2 representation. We define
\begin{equation}\label{Eq_g}
	g_{\alpha\beta}(\omega)=\frac{1}{2}
	\varepsilon_{\gamma\delta\alpha}\eta_{\gamma\delta\beta}(\omega),
\end{equation} 
where $g_{\alpha\beta}(\omega)$ 
is commonly known as the gyration tensor.
Crystal symmetries restrict the form of $g_{\alpha\beta}$,
making it trivially vanish 
in centrosymmetric crystals
\cite{10.1063/1.433207}.
In particular, 
among the 32 crystal classes, only 18 exhibit a nonzero
$g_{\alpha\beta}$ tensor as defined by 
Eq. (\ref{Eq_g}). By focusing on crystals with 3-fold point group symmetry, which includes the ferroelectric phase of PGO, the gyration tensor takes the following form,
\begin{equation}\label{Eq_g_tensor}
	g_{\alpha\beta}(\omega) =
	\begin{pmatrix}
		g_{11}(\omega) &  g_{12}(\omega) & 0      \\
		-g_{12}(\omega) &  g_{11}(\omega) & 0      \\
		0      &  0      & g_{33}(\omega)
	\end{pmatrix},
\end{equation}
where the optic axis is assumed to be parallel to the $z$ Cartesian 
direction. 
Note that the symmetric ($g_{33}$ and $g_{11}$) and antisymmetric ($g_{12}$)
components of the gyration tensor are associated to 
different observable physical effects.
The antisymmetric components 
$g_{12}=-g_{21}$ (corresponding to $\eta_{131}=\eta_{232}=-\eta_{311}=-\eta_{322}$)
are related to longitudinal excitons~\cite{agranovich1984crystal,IVCHENKO1978345,10.1063/1.433207} and describe longitudinal effects of NOA. 
For example, for a linearly polarized light along the
optic axis (which is assumed to be parallel to the 
$z$ Cartesian direction in our setting) propagating along the $x$ Cartesian direction, 
the $\eta_{131}$ component of the NOA gives rise to a longitudinal $E_1$ component of the electric field, since $\mathbf{E}\parallel \mathbf{q}$.
On the other hand, it is well known that
for an arbitrary direction of propagation only
the symmetric part of the gyration tensor
contributes to optical rotation
\cite{10.1063/1.433207,agranovich1984crystal}. In particular, 
the optical rotatory power for 
light propagating parallel to the optic (trigonal) axis 
depends exclusively on $g_{33}$,
\footnote{Moreover, unless light propagates parallel to the optic axis, optical rotation
acquires contributions from both the NOA and
the birefringence \cite{agranovich1984crystal}. 
Because birefringence is typically orders of magnitude stronger than NOA, isolating the latter is challenging. 
Therefore, optical activity measurements in PGO are generally performed with light propagating along the optic axis, where birefringence vanishes, allowing for a direct measurement 
of the $g_{33}$ tensor 
component \cite{10.1063/1.1661044,adamenko2023critical}.}
\begin{equation}
	\rho(\omega)=\frac{\omega^2}{2c^2}g_{33}(\omega),
\end{equation}
where $c$ is the speed of light and the gyration coefficients are
real, as we assume no dissipation \cite{agranovich1984crystal}. Following 
Ref. \cite{PhysRevLett.131.086902}, we focus
on the zero frequency limit, where the quantity $\bar{\rho}(\omega)=\rho(\omega)/(\hbar\omega)^2$ tends to a 
constant as $\omega\rightarrow 0$. Consequently,
\begin{equation}\label{Eq_bar_rho}
	\bar{\rho}=\frac{g_{33}}{2(\hbar c)^2},
\end{equation}
where $\hbar$ is the reduced Planck constant and 
\begin{equation}
	\bar{\rho}=\bar{\rho}(\omega\rightarrow 0), \quad
	g_{\alpha\beta}=g_{\alpha\beta}(\omega\rightarrow 0).
\end{equation}
Similarly, from now on we will assume that 
$\eta_{\alpha\beta\gamma}$ refers to its $\omega\rightarrow 0$
limit.

\subsection{NOA from DFPT}
In this section, we recall the main equations from Ref. \cite{PhysRevLett.131.086902}
that describe the implementation of natural optical activity as currently available in \textsc{abinit} \cite{GONZE2020107042,GONZE20092582}.
The NOA tensor is recast as a third-order derivative of the total energy,
\begin{equation}\label{Eq_eta}
	\eta_{\alpha\beta\gamma}= -\frac{4\pi}{\Omega}\text{Im }
	E_{\gamma}^{\mathcal{E}_\alpha\mathcal{E}_\beta},
\end{equation}
where $\Omega$ is the volume of the unit cell and 
\begin{equation}
		E_{\gamma}^{\mathcal{E}_\alpha\mathcal{E}_\beta}=
	\frac{\partial}{\partial q_\gamma}\left(
	\frac{d^2 E}{d \mathcal{E}^{-\mathbf{q}}_\alpha 
		d \mathcal{E}^\mathbf{q}_\beta}
	\right)\Bigg|_\mathbf{q=0}.
\end{equation}
Here, $\boldsymbol{\mathcal{E}}^\mathbf{q}$ represents a spatially
modulated electric field as introduced in Ref. \cite{PhysRevX.9.021050}. 
This quantity enters the Hamiltonian via the vector potential, within an 
electromagnetic gauge where the scalar potential vanishes and 
$\boldsymbol{\mathcal{E}}^\mathbf{q}=-\partial_t \mathbf{A}^\mathbf{q}$.
By exploiting the ``$2n+1$'' theorem, \cite{PhysRevX.9.021050} the term $E_{\gamma}^{\mathcal{E}_\alpha\mathcal{E}_\beta}$ in
Eq. (\ref{Eq_eta}) can be written exclusively in terms of
response functions to uniform field perturbations as \cite{PhysRevLett.131.086902}
\begin{equation}\label{Eq_EEE_gamma}
	E_{\gamma}^{\mathcal{E}_\alpha\mathcal{E}_\beta}=
	E_{\text{elst},\gamma}^{\mathcal{E}_\alpha\mathcal{E}_\beta}+
	2\int_\text{BZ}[d^3k]
	E_{\mathbf{k},\gamma}^{\mathcal{E}_\alpha\mathcal{E}_\beta},
\end{equation}
where  
$[d^3k]=\Omega/(2\pi)^3\,d^3k$ is a 
shorthand notation for the Brillouin-zone (BZ) integration and 
\begin{equation}
	E_{\text{elst},\gamma}^{\mathcal{E}_\alpha\mathcal{E}_\beta}=\int_\Omega\int n^{\mathcal{E}_\alpha}K_\gamma(\mathbf{r},\mathbf{r}')
	n^{\mathcal{E}_\beta}\,d^3r\,d^3r'.
\end{equation}
Here, $n^{\mathcal{E}_\alpha}(\mathbf{r})$ is the first-order electron density 
response to an electric field $\mathcal{E}_\alpha$ and 
$K_\gamma(\mathbf{r},\mathbf{r}')$ represents the derivative in $q_\gamma$
of the Hartree and exchange-correlation kernel. The quantity that needs to be 
integrated over the BZ in Eq. (\ref{Eq_EEE_gamma}) is given by \cite{PhysRevLett.131.086902}
\begin{equation}\label{Eq_symbols}
	\begin{split}
		E_{\mathbf{k},\gamma}^{\mathcal{E}_\alpha\mathcal{E}_\beta}=&
		\mathcal{X}_\mathbf{k}^{\mathcal{E}_\alpha k_\gamma\mathcal{E}_\beta}
		+\mathcal{Y}_\mathbf{k}^{\mathcal{E}_\alpha\mathcal{E}_\beta k_\gamma}
		+\mathcal{Y}_\mathbf{k}^{k_\gamma\mathcal{E}_\alpha\mathcal{E}_\beta}\\
		&+\mathcal{W}_\mathbf{k}^{\alpha,\beta\gamma}
        +\left( \mathcal{W}_\mathbf{k}^{\beta,\alpha\gamma} \right)^*.
	\end{split}
\end{equation}
For three generic perturbations ---$\lambda_1$, $\lambda_2$ and $\lambda_3$---
the $\mathcal{X}^{\lambda_1\lambda_2\lambda_3}_\mathbf{k}$ and 
$\mathcal{Y}^{\lambda_1\lambda_2\lambda_3}_\mathbf{k}$ symbols introduced in the last equation
are given by
\begin{subequations}
	\begin{align}
		\mathcal{X}_\mathbf{k}^{\lambda_1\lambda_2\lambda_3}=&
		\sum_m f_{m\mathbf{k}}\bra*{u^{\lambda_1}_{m\mathbf{k}}}\hat{\mathcal{H}}_\mathbf{k}^{\lambda_2}
		\ket*{u^{\lambda_3}_{m\mathbf{k}}},\\
		\mathcal{Y}_\mathbf{k}^{\lambda_1\lambda_2\lambda_3}=&
		-\sum_{m,n}f_{m\mathbf{k}}\bra*{u^{\lambda_1}_{m\mathbf{k}}}
		\ket*{u^{\lambda_3}_{n\mathbf{k}}}\bra*{u^{(0)}_{n\mathbf{k}}}
		\hat{\mathcal{H}}_\mathbf{k}^{\lambda_2}\ket*{u^{(0)}_{m\mathbf{k}}},
	\end{align}
\end{subequations}
where
$f_{m\mathbf{k}}$ is the occupation function,
$\ket*{u^\lambda_{m\mathbf{k}}}$ are first-order wave functions and 
$\hat{\mathcal{H}}_\mathbf{k}^\lambda=
\hat{H}_\mathbf{k}^\lambda+\hat{V}^\lambda$. Here, $\hat{H}_\mathbf{k}^\lambda$ 
represents the external perturbation and $\hat{V}^\lambda$ is the first-order
self-consistent field (SCF) potential response \footnote{Note that $\hat{\mathcal{H}}_\mathbf{k}^{k_\gamma}=\hat{H}_\mathbf{k}^{k_\gamma}$
	and we can set 
	$\hat{\mathcal{H}}_\mathbf{k}^{\mathcal{E}_\alpha}=\hat{V}^{\mathcal{E}_\alpha}$
	in all of our equations, since $\hat{H}_\mathbf{k}^{\mathcal{E}_\alpha}$ 
	is a purely ``cross-gap'' operator.} to $\lambda$.
The remaining symbol in Eq. (\ref{Eq_symbols}), 
$\mathcal{W}_\mathbf{k}^{\alpha,\beta\gamma}$, is given by
\begin{equation}
    \mathcal{W}_\mathbf{k}^{\alpha,\beta\gamma}=
    \sum_m f_{m\mathbf{k}}\bra*{u^{\mathcal{E}_\alpha}_{m\mathbf{k}}}
    \ket*{i\, u^{A_\beta}_{m\mathbf{k},\gamma}},
\end{equation}
where $\ket*{u^{\mathcal{E}_\alpha}_{m\mathbf{k}}}$ is the 
first-order wave function response to a uniform electric field and 
$\ket*{u^{A_\beta}_{m\mathbf{k},\gamma}}$ represents the response to a 
vector potential at first-order in the wave vector $\mathbf{q}$ 
(see the Appendix for more details).
Following Ref. \cite{PhysRevLett.131.086902}, 
we  split the latter into a sum of symmetric ($\mathcal{S}$) and
antisymmetric ($\mathcal{A}$) contributions with respect to $\beta\leftrightarrow\gamma$ exchange as
\begin{equation}
    \mathcal{W}_\mathbf{k}^{\alpha,\beta\gamma}=\frac{1}{2}
    \left(
    \mathcal{S}_\mathbf{k}^{\alpha,\beta\gamma}
    +\mathcal{A}_\mathbf{k}^{\alpha,\beta\gamma}
    \right),
\end{equation}
where
\begin{subequations}
	\begin{align}
		\frac{1}{2}\mathcal{S}^{\alpha,\beta\gamma}_\mathbf{k}&=\frac{i}{2}\sum_m f_{m\mathbf{k}}
		\bra*{u^{\mathcal{E}_\alpha}_{m\mathbf{k}}}
		\ket*{\partial^2_{\beta\gamma}u^{(0)}_{m\mathbf{k}}},\\
		\frac{1}{2}\mathcal{A}^{\alpha,\beta\gamma}_\mathbf{k}&=
        \varepsilon_{\gamma\beta\delta}
        \sum_m f_{m\mathbf{k}}
		\bra*{u^{\mathcal{E}_\alpha}_{m\mathbf{k}}}
		\ket*{u^{B_\delta}_{m\mathbf{k}}},\label{Eq_A_orb}
	\end{align}
\end{subequations}
where
$\partial_{\beta\gamma}^2=\partial^2/\partial k_\beta\partial k_\gamma$ represents a second-derivative in
$\mathbf{k}$ space and 
$\ket*{u^{B_\delta}_{m\mathbf{k}}}$ is the first-order wave function response to a uniform magnetic field. The latter accounts only for the 
orbital part of the response, i.e., the $B_\delta$ field does not interact 
with spins (see the Appendix for more details).
In all of our equations, it should be understood that the band indices $m,n$ run only over the occupied (valence) manifold.

\subsection{Including spins}\label{Sec_NOA_SOC}
In order to include the spin degrees of freedom into the formalism,
we consider the following ground-state 
Pauli-Schr\"odinger Hamiltonian, 
\cite{YAFET19631,PhysRevLett.116.077201}
\begin{equation}
    H^{(0)}=\frac{\mathbf{p}^2}{2m_e}+V(\mathbf{r})
    +\frac{\hbar}{4m_e}\left( \nabla V(\mathbf{r})\cross
    \mathbf{p} \right)\cdot
    \boldsymbol{\sigma},
\end{equation}
where $\mathbf{p}$ is the momentum operator, $V(\mathbf{r})$ is the periodic crystal potential, $m_e$ is the electron mass and 
$\boldsymbol{\sigma}$ is the Pauli vector, 
$\boldsymbol{\sigma}=(\sigma_x, \sigma_y,\sigma_z)$ 
containing the Pauli matrices.
When considering the effect of an electromagnetic field 
described by a vector potential 
$\mathbf{A}(\mathbf{r},t)=\mathbf{A}(\mathbf{q},\omega)e^{i(\mathbf{q\cdot r}-\omega t)}$, 
the latter equation becomes~\cite{YAFET19631}: 
\begin{equation}
\begin{split}
H=&\frac{1}{2m_e}(\mathbf{p}+e\mathbf{A})^2+V(\mathbf{r})\\
&+\frac{\hbar}{4m_e}\left[\nabla V(\mathbf{r})\cross (\mathbf{p}+e\mathbf{A}) \right]
\cdot\boldsymbol{\sigma}\\
&+g_s\frac{\mu_B}{2}(\nabla\cross \mathbf{A})\cdot\boldsymbol{\sigma},
\end{split}
\end{equation}
where $-e$ is the electron charge, $\mu_B=e\hbar/2m_e$ is the Bohr magneton
and $g_s\simeq 2.0023$ is the electron spin $g$ factor. We only keep terms
linear in the vector potential, 
$\hat{H}=\hat{H}^{(0)}+\hat{H}^{(1)}
+\mathcal{O}(\mathbf{A}^2)$, where $\hat{H}^{(0)}$ is the SOC-included ground-state Hamiltonian and
\begin{equation}\label{Eq_H1}
    \hat{H}^{(1)}=\hat{H}^\mathbf{A}_\text{orb}+\hat{H}_\text{spin}^\mathbf{A},
\end{equation}
where
$\hat{H}^\mathbf{A}_\text{orb}$ denotes the first-order Hamiltonian describing the coupling of the vector potential to the orbital degrees of freedom, as defined by Essin \textit{et al.} \cite{PhysRevB.81.205104}. 
The interaction between the spins and the vector potential is
treated exclusively by the second term on the right-hand side of 
Eq. (\ref{Eq_H1}),
\begin{equation}
\hat{H}_\text{spin}^{A_\alpha}= 
\frac{\partial}{\partial A_\alpha}
\left[\frac{g_s e}{2m_e}(\nabla\cross \mathbf{A})\cdot\mathbf{S}\right],
\end{equation}
where $\mathbf{S}$ is the spin operator, 
$\mathbf{S}=\frac{\hbar}{2}\boldsymbol{\sigma}$.
Assuming Hartree atomic units ($e=\hbar=m_e=1$) and 
$g_s\approx 2$ for simplicity, 
\begin{equation}\label{Eq_H_spin}
\hat{H}_\text{spin}^{A_c}=
i\,\varepsilon_{abc}\,q_a\,\hat{H}^{\mathfrak{B}_b},
\end{equation}
where $\hat{H}^{\mathfrak{B}_b}$  represents the 
first-order external perturbation associated to the spin contribution to a magnetic field, indicated with the symbol $\mathfrak{B}$, 
\begin{equation}
    \hat{H}^{\mathfrak{B}_\alpha}=-\frac{1}{2}\hat{\sigma}_\alpha.
\end{equation}
In the context of the equations derived in the previous section, 
the spin contribution coming from Eq. (\ref{Eq_H_spin}), which is linear
in the wave vector $\mathbf{q}$,
is relevant only for 
the response to a gradient of the vector potential, 
appearing in the $\mathcal{A}$ terms. 
It is straightforward to verify that, once the spin degrees of freedom are included, we can still write 
\begin{equation}\label{Eq_Berry}
\frac{1}{2}\mathcal{A}_\mathbf{k}^{\alpha,\beta\gamma}=
\varepsilon_{\gamma\beta\delta}
\sum_m f_{m\mathbf{k}}\bra*{u^{\mathcal{E}_\alpha}_{m\mathbf{k}}}
\ket*{u^{\mathcal{B}_\delta}_{m\mathbf{k}}}.
\end{equation}
This shows that the $\mathcal{A}$ contribution can again be expressed as a Berry curvature in the parameter space spanned by an electric field and a magnetic field, $\mathcal{B}$. However, $\mathcal{B}$ now accounts for both spin ($\mathfrak{B}$) and orbital ($B$) contributions.
(More details can be found in the Appendix).
It is implicitly assumed that the bras and kets
represent two component spinor wave functions and the
operators are expressed in spinorial form; that is, they are 
represented as $2\times 2$ matrices (equivalently, as four-vectors) \cite{PhysRevB.99.184404}.
\subsection{Treatment of spin-orbit coupling}
The only remaining task is to include
spin–orbit coupling in the calculation of all the individual
contributions to the NOA.
This entails introducing SOC into both the ground-state calculation and the response functions. 
Specifically, the latter requires the first and second derivatives with respect to the wave vector $\mathbf{k}$, 
and the response functions to a uniform electric and magnetic fields.
Spin-orbit coupling is already accounted for in the calculation of ground-state properties, the first derivatives with respect to $\mathbf{k}$, the 
response functions to an electric field and the spin part of the magnetic field response.
Furthermore,
the response function to the orbital part of a
magnetic field depends solely on first derivatives with respect to $\mathbf{k}$ 
(see Appendix),
which makes its extension to include SOC a straightforward task.
Hence, we have implemented this new option in \textsc{abinit}. This means that,
among all the ingredients required for the calculation of the NOA 
with SOC, only 
the $\ket*{\partial^2_{\beta\gamma}u^{(0)}_{m\mathbf{k}}}$ wave functions appearing in the $\mathcal{S}$ terms would require additional
coding effort.

While implementing this term with SOC is possible, we propose the following alternative strategy that allows us to avoid computing the $\mathcal{S}$ 
contributions entirely. As discussed in Ref. \cite{PhysRevLett.131.086902}, the analytical expression for $\eta_{\alpha\beta\gamma}$ is not unique. In fact, two different definitions of 
$E^{\mathcal{E}_\alpha\mathcal{E}_\beta}_{\mathbf{k},\gamma}$, as it appears in Eq. (\ref{Eq_EEE_gamma}), 
can yield identical gyration coefficients, provided their difference integrates to zero over the BZ. In light of this, we choose
\begin{equation}\label{Eq_E_tilde}
	\tilde{E}_{\mathbf{k},\gamma}^{\mathcal{E}_\alpha\mathcal{E}_\beta}
	= E_{\mathbf{k},\gamma}^{\mathcal{E}_\alpha\mathcal{E}_\beta} 
	- \frac{1}{2}f(\mathbf{k}),
\end{equation} 
where
\begin{equation}
		f(\mathbf{k})=
	\frac{\partial E_\mathbf{k,q}^{\mathcal{E}_\alpha\mathcal{E}_\beta}}
	{\partial k_\gamma} \bigg|_\mathbf{q=0},\quad \text{with }
	\int_\text{BZ}[d^3k]f(\mathbf{k})=0.
\end{equation}
Explicitly,
\begin{equation}
	\begin{split}
		f(\mathbf{k})=&
		\mathcal{X}_\mathbf{k}^{\mathcal{E}_\alpha\mathcal{E}_\beta k_\gamma}
		+\mathcal{X}_\mathbf{k}^{k_\gamma\mathcal{E}_\alpha\mathcal{E}_\beta}
		+\mathcal{X}_\mathbf{k}^{\mathcal{E}_\alpha k_\gamma \mathcal{E}_\beta}
		+\mathcal{Y}_\mathbf{k}^{\mathcal{E}_\alpha\mathcal{E}_\beta k_\gamma}\\
		&+\mathcal{Y}_\mathbf{k}^{\mathcal{E}_\alpha k_\gamma \mathcal{E}_\beta}
		+\mathcal{S}_\mathbf{k}^{\alpha,\beta\gamma}
		+\left(\mathcal{S}_\mathbf{k}^{\beta,\alpha\gamma}\right)^* .
	\end{split}
\end{equation}
According to Eq. (\ref{Eq_E_tilde}), 
the new (tilded) expression for the $\mathbf{k}$-dependent part
of the NOA is
\begin{equation}\label{Eq_E_k_gamma_prime}
	\begin{split}
		\tilde{E}_{\mathbf{k},\gamma}^{\mathcal{E}_\alpha\mathcal{E}_\beta}=
		\frac{1}{2}\Bigg[&
			\mathcal{X}_\mathbf{k}^{\mathcal{E}_\alpha k_\gamma \mathcal{E}_\beta}
			+\mathcal{Y}_\mathbf{k}^{\mathcal{E}_\alpha\mathcal{E}_\beta k_\gamma}
			+\mathcal{Y}_\mathbf{k}^{k_\gamma\mathcal{E}_\alpha\mathcal{E}_\beta}
            \\
		&-
        \left(
			\mathcal{X}_\mathbf{k}^{\mathcal{E}_\alpha\mathcal{E}_\beta k_\gamma}
			+\mathcal{X}_\mathbf{k}^{k_\gamma\mathcal{E}_\alpha\mathcal{E}_\beta}
			+\mathcal{Y}_\mathbf{k}^{\mathcal{E}_\alpha k_\gamma\mathcal{E}_\beta}
			\right)\\
		&+\mathcal{A}_\mathbf{k}^{\alpha,\beta\gamma}
			+\left(\mathcal{A}_\mathbf{k}^{\beta,\alpha\gamma}\right)^*
            \Bigg],
	\end{split}    
\end{equation}
where the $\mathcal{S}$ contributions do not appear.
In addition to treating the SOC case,
this expression significantly reduces the 
required number of linear-response calculations. 
Indeed, previously the computation of the NOA involved a total of 15 response functions: 3 from derivatives with respect to $\mathbf{k}$, 3 from derivatives with respect to an electric field, 
6 from second derivatives in  $\mathbf{k}$ and 3 from derivatives with respect to a magnetic field. 
Within our new approach, only 9 response functions are needed, as second derivatives 
in $\mathbf{k}$ are not required~\footnote{ 
Note, however, that the magnetic field perturbation now
acts on both the orbital and spin degrees of freedom.}.
%

\section{Results}\label{Sec_results}
In this section, we compute the natural optical activity of 
Pb$_5$Ge$_3$O$_{11}$ using our new implementation. We study the evolution of the NOA as a  
function of the ferroelectric polar distortion, and    
discuss the relevance of SOC effects, both directly,
through purely electronic effects, and
indirectly, via structural relaxation. In Sec. \ref{Sec_spin} we
focus entirely on the spin $\mathbf{B}$-field contributions to the Berry curvature
in Eq. (\ref{Eq_Berry}).
\subsection{Computational details}\label{Sec_details}
Our first-principles calculations are performed using the DFT and DFPT implementations of the
\textsc{abinit} package~\cite{GONZE2020107042,GONZE20092582}. 
We employ fully-relativistic norm-conserving pseudopotentials from
the PseudoDojo website~\cite{VANSETTEN201839},
under the PBEsol parametrization \cite{PhysRevLett.100.136406} of the 
Generalized Gradient Approximation (GGA).
A plane-wave energy cutoff of 40 Ha is applied and the Brillouin zone
is sampled using $3\times 3\times 3$ $\mathbf{k}$ points~\cite{fava2024}.
No significant changes on natural optical activity are observed 
when the $\mathbf{k}$-point mesh is increased up to 
$6\times 6\times 6$ $\mathbf{k}$ points.
Structures are relaxed to mechanical equilibrium
until the forces are smaller than $10^{-6}$ Ha/bohr.

Further methodological considerations 
arise when allowing a spin-magnetization response to a magnetic field
perturbation, 
since it requires the treatment
of a spin-density matrix in the calculations (see Sec. B of the Appendix for more details).
The DFPT routines that treat such cases in \textsc{abinit} 
are implemented \cite{PhysRevB.99.184404} within
the local density approximation (LDA) only, which is known to predict
lattice constants for PGO that deviate considerably more than those
obtained with the PBEsol functional \cite{fava2024}.
In addition, considering such a spin-density matrix increases memory usage 
and runtime by
roughly a factor of four, making our calculations along the full distortion 
path (see Sec. \ref{Sec_NOA_vs_dist}) infeasible with our available computational resources.
Under these circumstances, in Sec. \ref{Sec_NOA_vs_dist} and \ref{Sec_full_relax} 
we decided to stick to 
the PBEsol functional with SOC,
as it predicts lattice constants in better agreement with experiment, while 
acknowledging that the spin $\mathbf{B}$-field contributions in the $\mathcal{A}$ 
terms of Eq. (\ref{Eq_Berry}) are formally neglected in this setting.
The discussion of the spin magnetic field contributions is deferred to Sec. \ref{Sec_spin}.
\subsection{Natural optical activity as a function of the ferroelectric distortion}\label{Sec_NOA_vs_dist}
In order to study the evolution of the NOA
as a function of the ferroelectric distortion in PGO, we perform a linear interpolation, at fixed unit cell parameters, between the high-symmetry paraelectric (PE) phase and the
low-symmetry ferroelectric (FE) phase, which belong to the space group $P\bar{6}$ and $P3$, respectively.
We start by fully relaxing the PE phase
to mechanical equilibrium.
The obtained optimized lattice parameters and unit cell volume
are shown in Table \ref{Tab_strc_PE}. 
These values are in good agreement with experiments
\cite{doi:10.1143/JPSJ.43.961} and 
previous theoretical calculations \cite{fava2024}.
\begin{table}[t!]
	\caption{
		Lattice parameters $a$ and $c$ (in \AA) and unit cell volume $\Omega$
		(in \AA$^3$)  
        of the $P\bar{6}$ phase of PGO, after
		full optimization of the structure, with (w) and without (wo) SOC.
        The last row shows the experimental values reported in Ref. \cite{doi:10.1143/JPSJ.43.961}, at $473$ K. }
\label{Tab_strc_PE}
\begin{ruledtabular}  
\begin{tabular}{lccc}
	& $a$ & $c$ & $\Omega$\\
	\hline
    w SOC   & $10.221$  & $10.669$ & $965.267$ \\
	wo SOC & $10.214$  & $10.682$  & $965.083$ \\
    Exp. \cite{doi:10.1143/JPSJ.43.961} &$10.260$ & $10.696$& $\cdots$\\
    %
\end{tabular}
\end{ruledtabular} 
\end{table}
The reference FE phase that we use to make the linear interpolation is
obtained with the cell parameters fixed to those of the fully relaxed PE phase, while the ionic positions are relaxed. 
We define the amplitude of the global distortion as 
\begin{equation}\label{Eq_d}
	d=\sqrt{\frac{1}{M} \sum_\kappa  m_\kappa \abs{\mathbf{u}_\kappa}^2},
\end{equation}
where $m_\kappa$ is the mass of atom $\kappa$, $M = \sum_\kappa m_\kappa$ is the total mass of the unit cell, and $\mathbf{u}_{\kappa}$ denotes
the displacement of atom $\kappa$ 
from its position in the PE reference structure to its position in the FE phase. 

To assess the impact of spin-orbit coupling, we study three cases: (i) the case where SOC is completely neglected, (ii) the case where SOC is fully taken into account, 
including SOC-induced structural relaxation effects, and (iii) a case where 
the calculations are done without SOC, but 
with a structure frozen to the one obtained with SOC. 
The latter will enable us to account for SOC effects coming from the electrons only. 
This is of special relevance in the context of natural optical activity, where the structural parameters have
been shown to significantly alter the final result on their own~\cite{PhysRevLett.131.086902}.

\begin{figure}[t!]
	\includegraphics[width=1\linewidth]{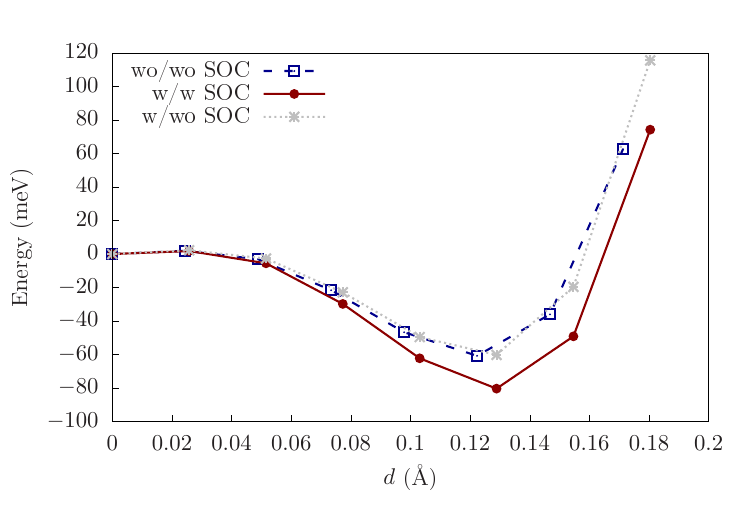}
	\caption{Evolution of the energy as a function of the FE distortion
		connecting the $P\bar{6}$ (at zero distortion) and $P3$ phases, where
		the reference energy corresponds to that of the PE phase.  
		The notation ``w/wo SOC'' indicates whether SOC is included in the structure relaxation (``w'' for with, ``wo'' for without) and in the subsequent DFT and DFPT calculations (e.g., energy or NOA). For example,
		``w/wo SOC'' means that the structure was relaxed with SOC, but calculations are performed without SOC. Solid and dashed lines are a guide 
		to the eye.} 
	\label{Fig_E}
\end{figure}

As a first step,  we show in Fig.~\ref{Fig_E} the evolution of the energy as a function of the amplitude of the global distortion $d$, as defined in Eq. (\ref{Eq_d}).
When SOC is taken into account for both the structural relaxation and the energy calculation (red bullets in Fig.~\ref{Fig_E}), we obtain a double-well 
depth
of $\Delta E=-80$ meV at a distortion amplitude $d=0.129$ \AA.
The
double-well energy depth is substantially reduced to $\Delta E=-61$ meV 
when SOC is not taken into account, neither during the relaxation nor in the energy calculation
(blue squares in Fig. \ref{Fig_E}), which is in agreement with previous works~\cite{fava2024}. 
 In this case, the minimum energy is achieved at a distortion amplitude of $d=0.122$ \AA.
%
Within the frozen SOC-optimized structure but removing SOC in the computation of the energy yields $\Delta E=-60$ meV 
(data in gray in Fig. \ref{Fig_E}). 
This value is very close to the case where the structure is 
relaxed without SOC, suggesting that the SOC-driven relaxation of the atomic
positions play only a minor role in the resulting double-well depth.
These substantial SOC effects on the potential energy landscape of lead germanate have been previously discussed in Ref. \cite{PhysRevB.108.L201112}. 

Next, by means of our new NOA implementation introduced in Sec. \ref{Sec_NOA_SOC},
we compute the evolution of the gyration coefficients as a function of the FE distortion.
\begin{figure}[b!]
	\includegraphics[width=1\linewidth]{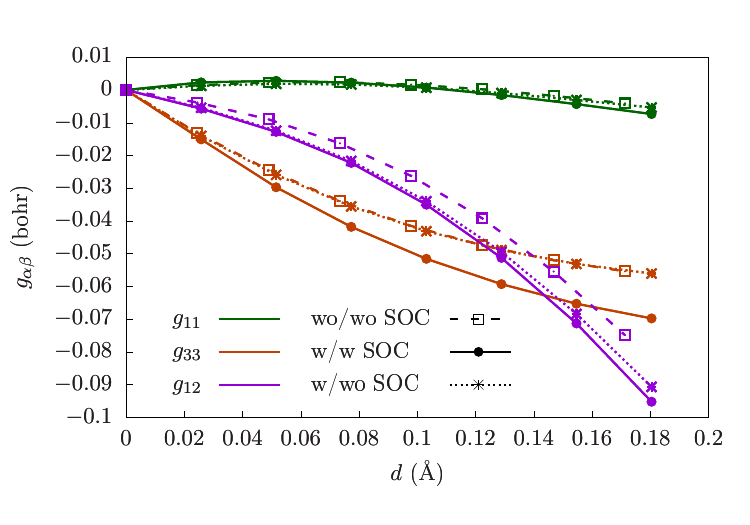}
	\caption{Evolution of the independent components of the gyration tensor ($g_{11}$, $g_{12}$, and $g_{33}$) as a function of the FE distortion
	connecting the $P\bar{6}$ (at zero distortion) and $P3$ phases.  
	Each individual component of the gyration tensor is represented with a different color, whereas the line style indicates
	whether SOC is included or not.
	(See the caption of Fig. \ref{Fig_E} for details on
	the notation.)
	Solid and dashed lines are a guide 
	to the eye.}
	\label{Fig_NOA}
\end{figure}
As shown in Fig.~\ref{Fig_NOA}, each  gyration component exhibits a different trend with increasing distortion. 
The absolute value of 
$g_{33}$ increases more rapidly than that of $g_{12}$ for
small distortions.
However, this behavior reverses for larger values
of $d$, where $|g_{12}|$ becomes larger than $|g_{33}|$ at roughly 
$d\sim 0.145 \text{ \AA}$  and  $d\sim 0.14  \text{ \AA}$, for the w/w SOC and wo/wo SOC cases, respectively. 
The physical reason of this behavior is difficult to identify in view of the complexity of the PGO structure and the FE distortions.

On the other hand, the $g_{11}$ component is small in magnitude and shows little variation across the whole distortion range. 
In spite of that, we observe an interesting behavior: it is positive up to a distortion of roughly 
$d\sim 0.1$ \AA, after which it becomes negative.  
\footnote{Recall that the distortion path under 
consideration does not correspond to the system's minimum energy path, as we are linearly interpolating between the initial (PE) and final (FE) configurations. 
As a consequence, the observed sign reversal might be an artifact of the chosen path, instead of a physically meaningful phenomenon.} 

Looking now at the impact of SOC on the gyration coefficients, we observe significant effects on both $g_{12}$ and $g_{33}$. 
Overall, the inclusion of SOC 
amplifies the absolute value of all the gyration components. 
Both structural and purely electronic contributions appear to play a comparable role in determining the final numerical value of $g_{12}$. 
In contrast, our results indicate that the SOC-induced contributions in $g_{33}$ arise predominantly from purely electronic effects.
This can be clearly deduced from Fig.~\ref{Fig_NOA}, where
the solid line (w/w SOC) and the dotted pattern line (w/wo SOC) are obtained with the same structure, and differ only 
from the fact that SOC has been activated or deactivated for the NOA calculation. 
The conclusions drawn for $g_{33}$ can be straightforwardly extended to the optical rotatory power, as defined by Eq. (\ref{Eq_bar_rho}).

\begin{figure}[b!]
	\includegraphics[width=1\linewidth]{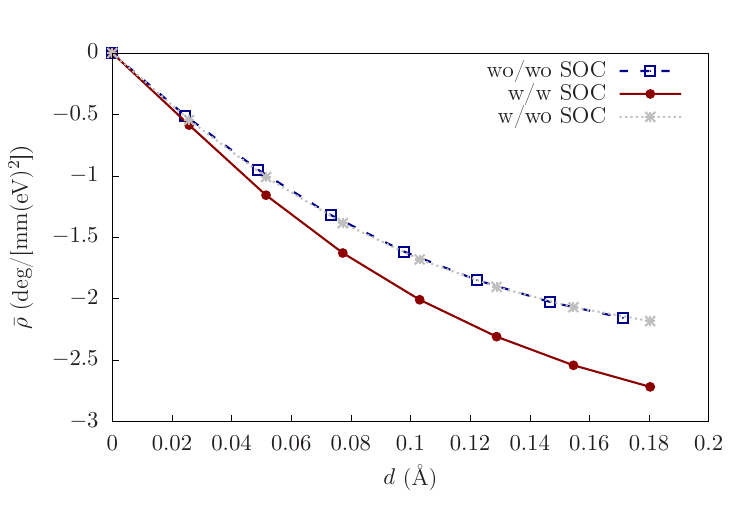}
	\caption{Evolution of the optical rotatory power of PGO
		as a function of the FE distortion
		connecting the $P\bar{6}$ (at zero distortion) and $P3$ phases.  
		(See the caption of Fig. \ref{Fig_E} for details on
		the notation.)
		Solid and dashed lines are a guide 
		to the eye.}
	\label{Fig_OR}
\end{figure}

We show in Fig.~\ref{Fig_OR} the optical rotatory power of PGO (in units of deg/[mm(eV)$^2$]) as a function of the FE distortion.
Unless SOC is explicitly included into the calculation of the NOA,
we can see that the SOC-induced relaxation effects have a little impact on the optical rotatory power.
Actually, when SOC is disabled in the NOA calculation, we obtain nearly identical results regardless of whether the structures used for interpolation are relaxed with or without SOC, which correspond, respectively, to the gray and blue 
data in Fig.~\ref{Fig_OR}. 
This suggests that the renormalization of atomic positions due to SOC plays only a minor role in determining the
magnitude of the optical rotation, and it indicates that the effect is primarily electronic in origin.
%

\subsection{Effect of full relaxation}\label{Sec_full_relax}
In this section, we present the calculated values for the independent components of the gyration tensor for the fully relaxed $P3$ phase of PGO, both with and without SOC.
By directly comparing these results with those from the previous section,
where the study was conducted at fixed cell parameters of the 
$P\bar{6}$ phase, 
we can evaluate the impact of full cell relaxation on the NOA. 
The structural details and the 
energy difference with respect to the reference $P\bar{6}$ phase, 
denoted by $\Delta E$, are reported in Table~\ref{Tab_strc}. The computed spontaneous
polarization along the hexagonal
c-axis agrees well with the experimental low-temperature saturation value of 
Ref. \cite{Shaldin2005}.

For a meaningful comparison with
 the quantities of Sec. \ref{Sec_NOA_vs_dist}, 
 we consider 
 the cases at the local minima of the double-well potential
 in Fig. \ref{Fig_E}, corresponding to 
 $d=0.129 \text{ \AA}$ for w/w SOC and w/wo SOC cases, and $d=0.122 \text{ \AA}$
 for the wo/wo SOC case.  
We provide in Table~\ref{Tab_NOA} the relevant $g_{\alpha\beta}$ tensor entries
and the optical rotatory power, for different cases under study.
\begin{table}[t!]
    \caption{
    Lattice parameters (in in \AA) and unit cell volume
    (in \AA$^3$) of the $P3$ phase of PGO and 
    energy difference (in meV) with respect to the reference $P\bar{6}$ phase. $P_z$ represents the
    projection of the spontaneous polarization (in $\mu$C/cm$^2$) along the $z$ Cartesian direction, parallel with the trigonal
    C3 rotation axis.
    The ``full'' label indicates relaxation of both
    cell parameters and ionic positions.} 
    \label{Tab_strc}
    \begin{ruledtabular}  
        \begin{tabular}{l
                        S[table-format=2.3]
                        S[table-format=2.3]
                        S[table-format=3.3]
                        S[table-format=3.0]
                        S[table-format=1.1]}
             & {$a$} & {$c$} & {$\Omega$} & {$\Delta E$} & {$P_z$}\\
            \hline
            w SOC (full)  & 10.257 & 10.689 & 973.900 & -89 & 5.5\\
            wo SOC (full) & 10.245 & 10.698 & 972.542 & -68 & 4.6\\
            Exp. 
            & 10.251 {$^a$} & 10.685 {$^a$} & \multicolumn{1}{c}{$\cdots$} & \multicolumn{1}{c}{$\cdots$} & 5.0 {$^b$}
        \end{tabular}
    \end{ruledtabular} 
    {\raggedright\footnotesize{$^{a}$ Ref. \cite{Sugii1971CrystalGA}}\par}
    {\raggedright\footnotesize{$^{b}$ Ref. \cite{Shaldin2005}}\par}
\end{table}
\begin{table}[t!]
	\caption{Independent components of the $g_{\alpha\beta}$ tensor (in bohr)
		and optical rotatory power (in deg/[mm(eV)$^2$]) of the $P3$ phase of PGO, for different cases. 
		}
	\label{Tab_NOA}
	\begin{ruledtabular}  
		\begin{tabular}{lrrrr}
			& \multicolumn{1}{c}{$g_{11}$} & \multicolumn{1}{c}{$g_{12}$} & \multicolumn{1}{c}{$g_{33}$} & \multicolumn{1}{c}{$\bar{\rho}$}\\
			\hline
			w/w SOC (full) & $-0.0006$ & $-0.0601$ & $-0.0607$& $-2.36$\\
			w/w SOC   & $-0.0015$ &$-0.0513$ & $-0.0593$ & $-2.31$\\
			\noalign{\vskip 8pt}  
			wo/wo SOC (full) & $0.0005$ & $-0.0457$ & $-0.0480$& $-1.87$\\
			wo/wo SOC  & $0.0002$ & $-0.0392$& $-0.0475$ & $-1.85$ \\
			\noalign{\vskip 8pt}  
			w/wo SOC (full)  & $-0.0002$ & $-0.0576$ & $-0.0496$ & $-1.93$\\
			w/wo SOC & $-0.0010$ & $-0.0495$ & $-0.0489$ & $-1.90$\\
		\end{tabular}
	\end{ruledtabular}
\end{table}
The data in Table~\ref{Tab_NOA} is gathered into pairs, so that the comparison between 
the fully relaxed cases and the results at fixed cell parameters
of Sec. \ref{Sec_NOA_vs_dist} is easier. 
These values indicate that
the effect of the full cell relaxation on $g_{33}$ 
is minimal, though larger deviations are observed for $g_{12}$ and $g_{11}$.  
In particular,
the deviation of ``w/w SOC'' and ``wo/wo SOC (full)'' cases
with respect to the reference ``w/w SOC (full)'' case are
$2\%$ and $21\%$, respectively, for the optical rotatory power.
The optical rotation in the ``w/wo SOC (full)'' case, which provides insight into the purely electronic SOC effects, shows a deviation of $18\%$, further supporting the conclusion drawn in the previous section 
that SOC contributions to the optical rotatory power in PGO
are predominantly electronic in nature.
We also note that $g_{11}$ is negative (positive) whenever the structure relaxation is performed with (without) SOC, irrespective of whether spin-orbit coupling is included in the subsequent calculations for the NOA.
Nonetheless, its absolute value
is two orders of magnitude smaller than that of $g_{12}$ and $g_{33}$ in nearly all  cases.

\subsection{Spin $\mathbf{B}$-field contributions}\label{Sec_spin}
With the goal of capturing the spin contributions to the $\mathcal{A}$ terms in Eq. (\ref{Eq_Berry}) coming from a
$\mathbf{B}$-field response, here we present calculations using fully relativistic pseudopotentials under
the Perdew-Wang \cite{PhysRevB.45.13244} LDA exchange-correlation (XC) functional
for two representative crystal structures of PGO: the experimental structure at 293 K \cite{10.1063/1.1661044} 
and the fully relaxed PBEsol structure from the previous subsection, which 
we treat as a 
$T=0$ K approximation.

We show in Table \ref{Tab_spin} the obtained numerical results for both structures under different 
computational settings. In particular, we use either LDA or PBEsol XC functionals, 
include or neglect SOC in the calculations, 
and consider or omit the spin $\mathbf{B}$-field contributions when 
doing calculations under the LDA. \footnote{Experimental values at $T=0$ K in Table \ref{Tab_spin} were obtained by graphical 
extrapolation of the data in 
Refs. \cite{10.1063/1.1661044,adamenko2023critical}, 
and thus should be regarded as approximate 
references only.}
\begin{table}[t!]
    \caption{Optical rotatory power (in deg/[mm(eV)$^2$]) of the $P3$ phase of PGO
    for the experimental structure \cite{ivanov2022new} at $T = 293$ K,
    and the fully relaxed 
    PBEsol structure ($T\sim0$ K), for different computational settings: SOC including 
    spin-$\mathbf{B}$ 
    contributions (SOC w spins) or excluding them (SOC wo spins) in 
    Eq. (\ref{Eq_Berry}), and calculations without SOC (no SOC), using either 
    LDA or PBEsol XC functionals.}
    \label{Tab_spin}
    \begin{ruledtabular}
        \begin{tabular}{lcccc}
            & \multicolumn{2}{c}{$T\sim0$ K}
            & \multicolumn{2}{c}{$T=293$ K} \\
            \cline{2-3} \cline{4-5}
            \noalign{\vskip 4pt}  
             & \multicolumn{1}{c}{PBEsol} & \multicolumn{1}{c}{LDA}
            & \multicolumn{1}{c}{PBEsol} & \multicolumn{1}{c}{LDA} \\
            \noalign{\vskip 4pt}  
            \hline
           	SOC w spins  &  $\cdots$   &$-2.48$& $\cdots$  &   $-1.86$     \\
			SOC wo spins &   $-2.36$      &$-2.42$  &  $-1.74$     & $-1.81$       \\
			no SOC       &   $-1.93$      &$-1.95$ &  $-1.39$     & $-1.40$       \\
            \noalign{\vskip 8pt}  
            Exp.  &\multicolumn{2}{c}{$-2.6^a$, $-2.3^b$}&
            \multicolumn{2}{c}{$-1.45^a$,$-1.43^b$}\\
        \end{tabular}
    \end{ruledtabular}
    {\raggedright\footnotesize{$^{a}$ Ref. \cite{10.1063/1.1661044}}\par}
    {\raggedright\footnotesize{$^{b}$ Ref. \cite{adamenko2023critical}}\par}
\end{table}

For a given structure, and within the same level of approximation, 
we see very good agreement between our PBEsol and
LDA values for the optical rotatory power, 
suggesting that the results depend
marginally on the choice of the XC functional, consistent with previous reports \cite{PhysRevLett.131.086902}. 
The spin $\mathbf{B}$-field contributions 
to the optical rotatory power of lead germante amplify the rest of spin-orbit 
coupling effects, though
their effect is small compared to the total SOC-induced contributions.
Such weak SOC-induced spin contributions are not unique to this system and have also
been discussed, e.g., in the context of the kinetic magnetoelectric effect (KME) in Ref. \cite{PhysRevResearch.6.013251}.
Magnetic materials are expected to exhibit larger contributions to the optical rotatory power 
from these spin-$\mathbf{B}$ terms.
This enhancement is expected to arise both directly, via the coupling of the external 
spin-$\mathbf{B}$ perturbation through the Pauli matrices on the right-hand side of the Sternheimer equation, and indirectly, through SCF effects induced by both the spin and orbital contributions to the $\mathbf{B}$-field response (see the Appendix for details).

For the structure relaxed within PBEsol ($T\sim 0$ K), the inclusion of spin-orbit coupling
is essential to obtain good agreement with experimental data extrapolated to $T=0$ K.
In contrast, for calculations based on the experimental structure at $T=293$ K,
we find, somewhat unexpectedly, that neglecting SOC leads to closer agreement
with experiments.
We note that the experimental lattice parameters and internal coordinates reported by
Iwata \textit{et al.} \cite{10.1063/1.1661044,Kay01011975} show small but non-negligible deviations from the more recent structural
measurements of Ref. \cite{ivanov2022new}, which we use as the reference structure for the $T=293$ K calculations.
Given that natural optical activity is known to be sensitive to even small structural details,
these structural differences may partly explain the larger deviations observed when 
spin-orbit coupling is included in our calculations, 
which are otherwise expected to provide a more accurate description
of the optical activity coefficients than the non-SOC approximation.
%
%

\section{Conclusions}\label{conclusions}
We have incorporated spin-orbit coupling into the DFPT methodology 
for natural optical activity~\cite{PhysRevLett.131.086902} as 
implemented in \textsc{abinit}. 
Our new approach enables a more realistic and accurate 
first-principles study of materials such as Pb$_5$Ge$_3$O$_{11}$,
where SOC-induced contributions can be substantial due to the presence of heavy atoms.
We have utilized our new implementation to study the 
natural optical activity of Pb$_5$Ge$_3$O$_{11}$ 
as a function of its ferroelectric distortion
from $P\bar{6}$ to $P3$ space groups.
We have shown that the magnitude of the optical 
rotation increases monotonically with ferroelectric distortion, 
persisting well beyond the double-well minimum into
energetically inaccessible regions.
Stabilizing such large distortions through chemical or 
structural \cite{zk45-6lb2}
engineering is a promising avenue for future work.
Beyond renormalizing the ferroelectric double-well as reported in
Ref. \cite{PhysRevB.108.L201112}, our work shows that
SOC also plays a fundamental role in the optical activity of PGO, where 
the main SOC-induced effects are electronic in origin, while SOC-driven structural relaxation effects
are minor. The two studies together thus provide a consistent picture of PGO as a material in which the interplay of structural cavities \cite{PhysRevB.108.L201112}, heavy-element chemistry and chirality gives rise to a remarkably rich set of SOC-driven phenomena, from structural energetics to optical response.

Our results of Sec. \ref{Sec_spin} indicate that, 
even in nonmagnetic systems 
like lead germanate, 
the spin-response to an external magnetic-field
gives rise to non-negligible contributions to the NOA when SOC is taken into account. 
Magnetic materials appear to be a particularly favorable scenario in which these contributions 
become more significant. 
We anticipate that the methodological 
advancements presented in this work will be valuable for establishing a full DFPT-based framework 
for calculating natural optical activity in such systems.

\begin{acknowledgments}
A.Z. and E.B. acknowledge the Fonds de la
Recherche Scientifique (FNRS) for support, the PDR
project CHRYSALID No.40003544, the EOS Project No. 560400077525 that has received funding from the FWO and FRS-FNRS under the Belgian Excellence of Science (EOS) program and the Consortium
des Équipements de Calcul Intensif (CÉCI), funded by
the F.R.S.-FNRS under Grant No. 2.5020.11 and the
Tier-1 Lucia supercomputer of the Walloon Region, 
infrastructure funded by the Walloon Region under the
grant agreement No. 1910247. 
M.S. acknowledges support by the Spanish MCIU/AEI/10.13039/501100011033 
through grant PID2023-152710NB-I00, and through a Severo Ochoa Excellence award to ICMAB, Grant
CEX2023-001263-S.

\end{acknowledgments}


\onecolumngrid
\begin{appendices}
\section{Linear-response to a static magnetic field}\label{App_a}
\subsection{Bare response (no SCF)}
We start by considering the following Sternheimer equation
\cite{PhysRevX.9.021050}
to a static $(\omega=0)$ but spatially modulated vector potential, 
$\mathbf{A}(\mathbf{r})=\mathbf{A}(\mathbf{q})e^{i\mathbf{q\cdot r}}$,
\begin{equation}\label{Eq_A_Stern}
    \left( \hat{H}_\mathbf{k+q}^{(0)}+\nu\hat{P}_\mathbf{k+q}
    -\epsilon^{(0)}_{m\mathbf{k}} \right)\ket*{u^{A_\beta}_{m\mathbf{k,q}}}
    =-\hat{Q}_\mathbf{k+q}
    \left(\hat{H}_\mathbf{k,q}^{A_\beta}
    +i\,q_a\,\varepsilon_{\beta ab}
    \hat{H}^{\mathfrak{B}_b}
    \right)\ket*{u^{(0)}_{m\mathbf{k}}},
\end{equation}
where $\nu$ is a parameter with dimensions of energy that ensures that
the left-hand side of the equation
does not become singular \cite{RevModPhys.73.515},
$\hat{P}_\mathbf{k}=\sum_m \ket*{u^{(0)}_{m\mathbf{k}}}
\bra*{u^{(0)}_{m\mathbf{k}}}$ is the projector onto the occupied states 
($\hat{Q}_\mathbf{k}=1-\hat{P}_\mathbf{k}$), 
$\hat{H}^{\mathfrak{B}_b}=-\frac{1}{2}\hat{\sigma}_b$ is the
first-order Hamiltonian  
associated with a magnetic-field interacting with the spin degrees of freedom,
and $\hat{H}^{A_\beta}_\mathbf{k,q}$ is the momentum representation of the
first-order Hamiltonian (orbital part) to a vector potential 
\cite{PhysRevB.81.205104,PhysRevX.9.021050,PhysRevB.98.075153}. 
For our purposes, it is enough to recall that
\begin{equation}
    \hat{H}_{\mathbf{k},\mathbf{q=0}}^{A_\beta}=
    \partial_\beta\hat{H}_\mathbf{k}^{(0)},\qquad
    \frac{\partial \hat{H}_\mathbf{k,q}^{A_\beta}}{\partial q_\gamma}
    \Bigg|_\mathbf{q=0}=\frac{1}{2}
    \partial^2_{\beta\gamma}\hat{H}_\mathbf{k}^{(0)},
\end{equation}
where $\partial_\gamma\equiv \partial/\partial q_\gamma$ and
$\partial^2_{\beta\gamma}\equiv \partial^2/\partial q_\gamma\partial q_\beta$
represent first and second derivatives with respect to the wave vector.
For simplicity, SCF contributions are excluded from Eq. (\ref{Eq_A_Stern}) at this stage 
and will be included later once the magnetic-field perturbation is isolated.

Taking the first derivative of Eq. (\ref{Eq_A_Stern}) with respect to $q_\gamma$
yields an expression for the wave function response to a gradient of
the vector potential, which acquires contributions belonging both to the valence- and  
conduction-band manifolds (indicated with a bar), \cite{PhysRevX.9.021050}
\begin{equation}\label{Eq_bar}
    \ket*{u^{A_\beta}_{m\mathbf{k},\gamma}}=
    -\partial_\gamma\hat{P}_\mathbf{k}\partial_\beta\hat{P}_\mathbf{k}
    \ket*{u^{(0)}_{m\mathbf{k}}}+
    \ket*{\bar{u}^{A_\beta}_{m\mathbf{k},\gamma}},
\end{equation}
where $[\hat{A},\hat{B}]=\hat{A}\hat{B}-\hat{B}\hat{A}$ and $ \ket*{\bar{u}^{A_\beta}_{m\mathbf{k},\gamma}}$
is a linear-response quantity that
fulfills the following
Sternheimer equation, \cite{PhysRevX.9.021050}
\begin{equation}\label{Eq_Stern_A_grad}
\left( \hat{H}_\mathbf{k}^{(0)} +\nu\hat{P}_\mathbf{k} 
-\epsilon^{(0)}_{m\mathbf{k}} \right)\ket*{\bar{u}^{A_\beta}_{m\mathbf{k},\gamma}}
=-\hat{Q}_\mathbf{k}\left[\hat{O}_\mathbf{k}^{\beta\gamma}
+i\,\varepsilon_{\beta\gamma\delta}\hat{H}^{\mathfrak{B}_\delta} 
\right]
\ket*{u^{(0)}_{m\mathbf{k}}},
\end{equation}
where
\begin{equation}
    \ket*{\bar{u}^{A_\beta}_{m\mathbf{k},\gamma}}\equiv
    \frac{\partial \ket*{\bar{u}^{A_\beta}_{m\mathbf{k,q}}}}{\partial q_\gamma}
    \Bigg|_\mathbf{q=0}
\end{equation}
and 
\begin{equation}
\hat{O}_\mathbf{k}^{\beta\gamma}=
\partial_\gamma\hat{H}^{(0)}_\mathbf{k}\partial_\beta\hat{P}_\mathbf{k}
-\partial_\gamma\hat{P}_\mathbf{k}\partial_\beta\hat{H}^{(0)}_\mathbf{k}
+\frac{1}{2}\partial^2_{\beta\gamma}\hat{H}^{(0)}_\mathbf{k}.
\end{equation}
The magnetic-field response is captured by the antisymmetric
($\beta\leftrightarrow\gamma$) components of the response.
We define the perturbing operator appearing in 
Eq. (\ref{Eq_Stern_A_grad}) 
as 
\begin{equation}
    \hat{T}^{\beta\gamma}_\mathbf{k}=\hat{O}_\mathbf{k}^{\beta\gamma}
+i\,\varepsilon_{\beta\gamma\delta}\hat{H}^{\mathfrak{B}_\delta},
\end{equation}
where $\hat{T}^{\beta\gamma}_\mathbf{k}=\frac{1}{2}(
\hat{T}_\mathbf{k}^{\text{S},\beta\gamma}
+\hat{T}_\mathbf{k}^{\text{A},\beta\gamma})$.
The symmetric (S) and antisymmetric (A) perturbing operators are given 
by
\begin{equation}
\begin{split}
\hat{T}_\mathbf{k}^{\text{S},\beta\gamma}&=
\partial_\gamma\hat{H}^{(0)}_\mathbf{k}\partial_\beta\hat{P}_\mathbf{k}
-\partial_\gamma\hat{P}_\mathbf{k}\partial_\beta\hat{H}^{(0)}_\mathbf{k}
+\partial_\beta\hat{H}^{(0)}_\mathbf{k}\partial_\gamma\hat{P}_\mathbf{k}
-\partial_\beta\hat{P}_\mathbf{k}\partial_\gamma\hat{H}^{(0)}_\mathbf{k}
+\partial^2_{\beta\gamma}\hat{H}^{(0)}_\mathbf{k},\\
\hat{T}_\mathbf{k}^{\text{A},\beta\gamma} 
&=
\partial_\gamma\hat{H}^{(0)}_\mathbf{k}\partial_\beta\hat{P}_\mathbf{k}
-\partial_\gamma\hat{P}_\mathbf{k}\partial_\beta\hat{H}^{(0)}_\mathbf{k}
-\partial_\beta\hat{H}^{(0)}_\mathbf{k}\partial_\gamma\hat{P}_\mathbf{k}
+\partial_\beta\hat{P}_\mathbf{k}\partial_\gamma\hat{H}^{(0)}_\mathbf{k}
+2i\,\varepsilon_{\beta\gamma\delta}\hat{H}^{\mathfrak{B}_\delta}.
\end{split}
\end{equation}
We can split the solution of Eq. (\ref{Eq_Stern_A_grad}) into its
symmetric and antisymmetric components as
\begin{equation}
    \ket*{\bar{u}^{A_\beta}_{m\mathbf{k},\gamma}}=\frac{1}{2}
    \left( \ket*{\bar{u}^{\text{S},A_\beta}_{m\mathbf{k},\gamma}}
    +\ket*{\bar{u}^{\text{A},A_\beta}_{m\mathbf{k},\gamma}}
    \right).
\end{equation}
It is then easy to check that 
$\ket*{\bar{u}^{\text{S},A_\beta}_{m\mathbf{k},\gamma}}=
\ket*{\partial^2_{\beta\gamma}u^{(0)}_{m\mathbf{k}}}$ and 
\begin{equation}
\frac{i}{2}\ket*{\bar{u}^{\text{A},A_\beta}_{m\mathbf{k},\gamma}}=
-\varepsilon_{\beta\gamma\delta}\left( 
\ket*{\bar{u}^{B_\delta}_{m\mathbf{k}}}+\ket*{\bar{u}^{\mathfrak{B}_\delta}_{m\mathbf{k}}}
\right)
=\varepsilon_{\gamma\beta\delta}\ket*{\bar{u}^{\mathcal{B}_\delta}_{m\mathbf{k}}},
\end{equation}
where we have defined the perturbing operator for the orbital part 
of a magnetic-field as
\begin{equation}\label{Eq_B_orb}
    \hat{O}_\mathbf{k}^{B_c}=-\frac{i}{2}
    \varepsilon_{abc}
    \{\partial_a\hat{P}_\mathbf{k},\partial_b\hat{H}^{(0)}_\mathbf{k}\},
\end{equation}
where $\{\hat{A},\hat{B}\}=\hat{A}\hat{B}+\hat{B}\hat{A}$. The total magnetic-field response 
contains contributions from the valence-band manifold coming from the first term on the
right-hand side of Eq. (\ref{Eq_bar}), which play an important role once SCF contributions are included (see next subsection).
Note that the definition for the operator 
$\hat{O}^{B_c}_\mathbf{k}$ given by Eq. (\ref{Eq_B_orb}) differs by a minus sign from the convention 
adopted in Ref. \cite{PhysRevLett.131.086902}. This is why the 
order of the indices
in the Levi-Civita symbol is changed in the $\mathcal{A}$ terms
between Ref. \cite{PhysRevLett.131.086902} and the present work.
In addition, we have
identified a typographical error in Eq. (17) of 
Ref. \cite{PhysRevLett.131.086902}, 
where the half factor should not appear; this has been corrected in our 
Eq. (\ref{Eq_A_orb}) and Eq. (\ref{Eq_Berry}).
\subsection{Self-consistent response}\label{app_scf}
Self-consistency in the magnetic-field response is enforced by incorporating the first-order potential $\hat{V}^{B_c}$ into its corresponding Sternheimer equation. 
Writing down the operators explicitly,
\begin{equation}\label{Eq_Stern_B}
    \left(\hat{H}_\mathbf{k}^{(0)}+\nu\hat{P}_\mathbf{k}-\epsilon^{(0)}_{m\mathbf{k}}\right)
    \ket*{\bar{u}^{\mathcal{B}_c}_{m\mathbf{k}}}=-\hat{Q}_\mathbf{k}\left(
    -\frac{i}{2}\varepsilon_{abc}\{\partial_a\hat{P}_\mathbf{k},\partial_b\hat{H}^{(0)}_\mathbf{k}\}
    -\frac{1}{2}\hat{\sigma}_c+\hat{V}^{\mathcal{B}_c}
    \right)\ket*{u^{(0)}_{m\mathbf{k}}},
\end{equation}
where
\begin{equation}\label{Eq_V}
    V^{\mathcal{B}_c}(\mathbf{r})=\int K_\text{Hxc}(\mathbf{r,r'})n^{\mathcal{B}_c}(\mathbf{r})\,d^3r'.
\end{equation}
In the last equation, $K_\text{Hxc}(\mathbf{r,r'})$ is the Hartree and exchange-correlation (Hxc) kernel, 
which is a $4\times 4$ matrix.
The first-order electron density appearing in Eq. (\ref{Eq_V}) can be written as the following 
trace in spinor space, \cite{PhysRevB.99.184404,PRX_noncollinear}
\begin{equation}\label{Eq_n1}
    n^{\mathcal{B}_c}(\mathbf{r})=\frac{1}{2}\Tr\left[\boldsymbol{\sigma}\int_\text{BZ}[d^3k]\sum_m f_{m\mathbf{k}}\left(
    \bra*{\mathbf{r}}\ket*{u^{\mathcal{B}_c}_{m\mathbf{k}}}\bra*{u^{(0)}_{m\mathbf{k}}}\ket*{\mathbf{r}}
    +\text{h.c.}
    \right)\right].
\end{equation}
Here, ``h.c." denotes hermitian conjugation, and 
$\boldsymbol{\sigma}=(\sigma_0,\sigma_x,\sigma_y,\sigma_z)$ collects the $2\times2$ identity matrix $\sigma_0$ together with the Pauli matrices $\sigma_i$ ($i=x,y,z$). Similarly, 
$n^{\mathcal{B}_c}=(n_0^{\mathcal{B}_c},n_x^{\mathcal{B}_c},n_y^{\mathcal{B}_c},n_z^{\mathcal{B}_c})$ 
is a four-vector. To emphasize the role of the valence-band contributions in
the magnetic-field response discussed above, let us explicitly write the first-order wave function 
response as 
\begin{equation}
    \ket*{u_{m\mathbf{k}}^{\mathcal{B}_c}}=
    -i\,\varepsilon_{abc}\partial_a\hat{P}_\mathbf{k}\partial_b\hat{P}_\mathbf{k}\ket*{u^{(0)}_{m\mathbf{k}}}+
    \ket*{\bar{u}_{m\mathbf{k}}^{\mathcal{B}_c}}.
\end{equation}
Substituting this expression into Eq. (\ref{Eq_n1}) yields
\begin{equation}
\begin{split}
 n^{\mathcal{B}_c}(\mathbf{r})=&\frac{1}{2}\Tr\left[\boldsymbol{\sigma}\int_\text{BZ}[d^3k]\sum_m f_{m\mathbf{k}}\left(
  \bra*{\mathbf{r}}\ket*{\bar{u}^{\mathcal{B}_c}_{m\mathbf{k}}}
  \bra*{u^{(0)}_{m\mathbf{k}}}\ket*{\mathbf{r}}
    +\text{h.c.}
 \right)\right]\\
 &-\frac{i}{2}\,\varepsilon_{abc}\,\Tr\left[\boldsymbol{\sigma}\int_\text{BZ}[d^3k]\sum_{m,n} f_{m\mathbf{k}}\left(
  \bra*{\mathbf{r}}\ket*{u^{(0)}_{n\mathbf{k}}}
  \bra*{\partial_a u^{(0)}_{n\mathbf{k}}}\ket*{\partial_b u^{(0)}_{m\mathbf{k}}}
  \bra*{u^{(0)}_{m\mathbf{k}}}\ket*{\mathbf{r}}
 +\text{h.c.}\right)\right].
 \end{split}
\end{equation}
The first term is the usual DFPT density-response contribution, given by ground-state Bloch functions and 
their first-order (Sternheimer-like) corrections. The second term arises in the particular case of a 
magnetic-field perturbation, specifically its orbital part, and involves covariant $\mathbf{k}$-derivatives 
of Bloch states, yielding an orbital Berry-curvature-type contribution to the density-response.
Note that, in time-reversal symmetric insulators, the latter contribution is nonzero only in the spin 
components of the density response, $(n_x^{\mathcal{B}_c},n_y^{\mathcal{B}_c},n_z^{\mathcal{B}_c})$, 
while the charge component $n_0^{\mathcal{B}_c}$ vanishes identically.
In practice, and according to the implementation limitations discussed in Sec. \ref{Sec_details}, 
these additional 
contributions to the density-response and the induced first-order SCF potential 
---see Eq. (\ref{Eq_V})--- are currently accessible in \textsc{abinit} (v10.6.5) 
only within the LDA XC approximation.
\end{appendices}

\twocolumngrid
\bibliography{biblio}

\end{document}